\pdfoutput=1
\documentclass[%
superscriptaddress,
showpacs, preprintnumbers,
amsmath,amssymb,
aps,
prl,
twocolumn,
10pt,
]{revtex4-1}
\usepackage{booktabs}
\usepackage[para]{threeparttable}
\usepackage[utf8]{inputenc}
\usepackage{hyperref}
\usepackage{color}
\usepackage{units}
\usepackage{upgreek}

\usepackage{graphicx}
\usepackage{bm}
\usepackage{dcolumn}   

\begin{document}
\title{Tuning the van der Waals Interaction of Graphene with Molecules via Doping}
\author{Felix Huttmann}\email{huttmann@ph2.uni-koeln.de}
\affiliation{II. Physikalisches Institut, Universit\"{a}t zu K\"{o}ln, 
Z\"{u}lpicher Stra\ss e 77, 50937 K\"{o}ln, Germany}
\author{Antonio J. Martínez-Galera}
\affiliation{II. Physikalisches Institut, Universit\"{a}t zu K\"{o}ln, 
Z\"{u}lpicher Stra\ss e 77, 50937 K\"{o}ln, Germany}
\author{Vasile Caciuc}
\affiliation{Peter Grünberg Institute and Institute for Advanced Simulation, 
Forschungszentrum Jülich, 52428 Jülich, Germany}
\author{Nicolae Atodiresei}\email{n.atodiresei@fz-juelich.de}
\affiliation{Peter Grünberg Institute and Institute for Advanced Simulation, 
Forschungszentrum Jülich, 52428 Jülich, Germany}
\author{Stefan Schumacher}
\affiliation{II. Physikalisches Institut, Universit\"{a}t zu K\"{o}ln, 
Z\"{u}lpicher Stra\ss e 77, 50937 K\"{o}ln, Germany}
\author{Sebastian Standop}
\affiliation{II. Physikalisches Institut, Universit\"{a}t zu K\"{o}ln, 
Z\"{u}lpicher Stra\ss e 77, 50937 K\"{o}ln, Germany}
\author{Ikutaro Hamada}
\affiliation{International Center for Materials Nanoarchitectonics 
(WPI-MANA) and Global Research Center for Environment and Energy
based on Nanomaterials Science (GREEN), National Institute for Materials 
Science, 1-1 Namiki, Tsukuba 305-0044, Japan}
\author{Tim O. Wehling}
\affiliation{Bremen Center for Computational Material Science (BCCMS), 
Universität Bremen, Am Fallturm 1a, 28359 Bremen, Germany}
\author{Stefan Blügel}
\affiliation{Peter Grünberg Institute and Institute for Advanced Simulation, 
Forschungszentrum Jülich, 52428 Jülich, Germany}
\author{Thomas Michely}
\affiliation{II. Physikalisches Institut, Universit\"{a}t zu K\"{o}ln, 
Z\"{u}lpicher Stra\ss e 77, 50937 K\"{o}ln, Germany}
\date{\today}
\newcommand {\fh}[1]{{\color{magenta} #1}}
\newcommand {\tm}[1]{{\color{blue} #1}}
\begin{abstract}
We use scanning tunneling microscopy to visualize and thermal desorption spectroscopy to quantitatively measure  that the binding of naphthalene molecules to graphene (Gr), a case of pure van der Waals (vdW) interaction, strengthens with $n$- and weakens with $p$-doping of Gr.
Density functional theory calculations that include the vdW interaction in a seamless, \emph{ab initio} way accurately reproduce the observed trend in binding energies. Based on a model calculation, we propose that the vdW interaction is modified by changing the spatial extent of Gr's $\pi$ orbitals via doping.
\end{abstract}

\maketitle
One of the key properties of graphene (Gr) is the wide-range tunability of its Fermi level and corresponding charge carrier concentration, either by a gate electrode~\cite{Novoselov04}, substitutional doping \cite{ACSCatal.2.781}, adsorption \cite{Bostwick10, PhysRevB.81.235401}, or charge transfer from a supporting material or intercalation layer \cite{PhysRevLett.103.246804,Petrovic2013, ACSNano.6.9551, PhysRevB.90.235437}.
The tunability of the Fermi level through the otherwise rigid band structure results from the material being atomically thin and having a negligible density of states near the Dirac point.

In recent years, interest has arisen in using this tunability to control adsorption:
For the case of ionic adsorbates, Brar \emph{et al.} \cite{Brar10} demonstrated a dependence of the ionization state of a Co adatom on the Gr Fermi level position, and Schumacher \emph{et al.} \cite{doi:10.1021/nl402797j} found a doping-dependent binding energy $E_b$ of ionic adsorbates to Gr, with a shift in $E_b$ on the order of the shift in the Fermi level induced by doping. For the case of radicals and based on \emph{ab initio} calculations, Wehling \emph{et al.} \cite{PhysRevB.90.085422} predict doping-dependent adsorbate phase transitions for hydrogenated as well as fluorinated Gr, while Huang et al. \cite{Huang11} find a stronger binding of isolated H radicals for larger magnitudes of doping. 

For the case of van der Waals (vdW) interaction, the effect of the Gr doping level on the binding energy of adsorbates has not yet been explored.
This is surprising, given that the adsorption of simple hydrocarbons to graphite or Gr has been used as a model system to study vdW interactions~\cite{Charakova06, PhysRevB.69.155406,JPhysCondMat.24.424212,JChemPhys.125.234308}.
Here, we investigate this case with the help of epitaxial Gr on Ir(111), which can be doped from the backside by intercalation of highly electropositive (e. g. Cs, Eu) or \mbox{-negative} (e.g. O) elements into its interface with the substrate, while Gr's other side remains available for the adsorption experiment itself. This strategy not only enables us to achieve large Fermi level shifts on the order of $\pm 1$\,eV, but also to visualize doping-induced binding energy differences by making using of intercalation patterns \cite{doi:10.1021/nl402797j}. Naphthalene is chosen as a test molecule, since its binding to Gr is a pure vdW case studied previously, both experimentally \cite{PhysRevB.69.155406} and theoretically \cite{Charakova06}. Our experiments are complemented by density functional theory (DFT) calculations that include the vdW interaction in a seamless, \emph{ab initio} way 
(for a recent review, see Ref.\,\cite{RepProgPhys.78.066501}). 

For this paradigmatic case we find in excellent agreement of experiment and theory an increase of the vdW binding energy when changing from $p$- to $n$-doping. We identify the mechanism of this vdW binding energy manipulation, which is likely to hold for a broad class of molecular adsorbates and a variety of 2D materials.

The experiments were performed in two variable-temperature scanning tunneling microscopy (STM) systems with base pressures below $\unit[1 \times 10^{-10}]{mbar}$. Ir(111) was prepared by cycles of sputter-anneal, yielding clean terraces with sizes on the order of \unit[100]{nm}. Well oriented Gr was prepared by room temperature ethylene adsorption till saturation, thermal decomposition at \unit[1500]{K} and subsequent exposure to $\unit[5 \times 10^{-7}]{mbar}$ of ethylene at \unit[1170]{K} \cite{vanGastel09}. Using \unit[600]{s} ethylene exposure time, a perfectly closed, 1~monolayer~(ML) Gr layer was achieved, while with \unit[180]{s} a partial $\approx$ 0.9\,ML Gr layer resulted. For $n$-doping, Eu was intercalated under 1\,ML Gr on Ir(111) at 550\,K by exposure to high purity Eu \cite{ames} vapor.
For $p$-doping, oxygen was intercalated till saturation under 0.9\,ML Gr on Ir(111) at \unit[530]{K} using a local oxygen pressure of about $\unit[2.5\times 10^{-4}]{mbar}$ for \unit[600]{s}. 
Naphthalene exposure \footnote{Naphthalene ``analytical standard'' obtained from Sigma-Aldrich} was controlled by a variable leak valve.
STM images were taken at \unit[35]{K} and digitally post-processed with the WSxM software \cite{WSXM}.
Thermal desorption spectroscopy (TDS) was performed with a sample heating rate of $\unit[1]{Ks^{-1}}$ by monitoring desorbing naphthalene (molecular mass $m=\unit[128]{u}$).
The sample temperature measurement during TDS with a K-type thermocouple has an absolute error below \unit[3]{K} with a reproducibility better than \unit[1]{K}.

The DFT calculations were carried out in the generalized gradient
approximation (GGA) \cite{PRL77_3865} with a kinetic energy 
cut-off of 500\,eV using the projector augmented wave method (PAW) \cite{PRB50_17953} 
as implemented in the VASP code \cite{PRB47_558,PRB54_11169}.
The ground-state geometry and the corresponding electronic structure were investigated using the non-local correlation 
energy functional vdW-DF2 \cite{PRB82_081101} with a revised~\cite{PRB89_121103R} 
Becke (B86b) exchange energy functional \cite{JCP86_7184}. 
The unit cell comprised (10$\times$10) graphene unit cells on 4 layers of 
(9$\times$9) Ir(111), either pristine or (Eu/O)-intercalated.

\begin{figure}[b]
	\begin{center}
	\includegraphics{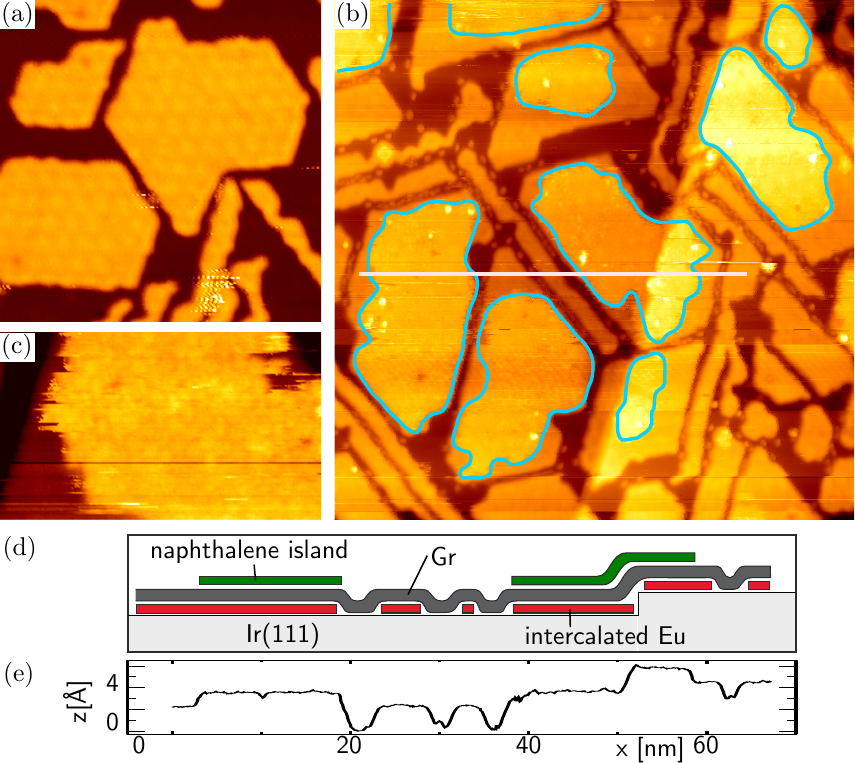}
	\end{center}
	\caption{\label{fig:stm}(color online)
		(a-c):~STM topographs imaged at $\unit[35]{K}$.
		(a):~Gr/Ir(111) after intercalation of about half 
		a monolayer coverage of Eu ($(\unit[39]{nm})^2$).
		(b):~Like (a), but with additional submonolayer coverage of naphthalene 
		($(\unit[90]{nm})^2$).
		(c):~Molecular resolution of naphthalene island ($\unit[(18\times 11)]{nm^2}$).
		(d):~Schematic cross section corresponding to profile in (e).
		(e):~Height profile along the white line indicated in (b).
	}
\end{figure}

We start with a discussion of the STM preparations and measurements. Upon intercalation of submonolayer amounts of Eu, a complex pattern of stripes and islands of Eu-intercalated Gr (high/bright) in coexistence with non-intercalated Gr (low/dark) emerges as shown in the STM topograph in Fig.~\ref{fig:stm}(a) and discussed in Ref.~\cite{PhysRevLett.110.086111}.
Within the intercalated patches, Eu forms a p$(2\times 2)$ superstructure and leads to local doping of graphene by $\unit[-1.36]{eV}$, while the surrounding pristine Gr areas remain hardly doped at \unit[+0.1]{eV} \cite{doi:10.1021/nl402797j}.

After deposition of submonolayer amounts of naphthalene at $T \unit[<50]{K}$ and subsequent brief annealing to $T=\unit[150]{K}$, the sample is cooled down again for STM imaging. The corresponding Fig.~\ref{fig:stm}\,(b) displays additional naphthalene islands (cyan encircled) only on top of Eu-intercalated patches. The profile in Fig.~\ref{fig:stm}\,(e) taken along the white line in Fig.~\ref{fig:stm}\,(b) shows that the apparent height
of the Eu intercalation island itself stays unchanged at \unit[2]{\AA}, while the naphthalene islands have an apparent height of about \unit[1.3\dots 1.5]{\AA}
on top of the Eu-intercalated Gr. The schematic cross section of Fig.~\ref{fig:stm}\,(d) visualizes the morphology underlying the height profile.
Higher-resolution imaging displayed in Fig.~\ref{fig:stm}\,(c) reveals molecular resolution of the naphthalene adsorbate layer.
Our interpretation is as follows: During annealing, the molecules are highly mobile, diffuse over the surface while sensing local adsorption energy differences. As the sample is cooled down again, the arrangement is frozen and 
the preferential coverage of intercalated areas by molecules then directly reflects the preferential binding to these areas.

To strengthen our case for the generality of this result, we repeated the experiment exchanging naphthalene with benzene, and found qualitatively the same results {(compare \cite{SM})}.

\begin{figure}
\includegraphics[width=\columnwidth]{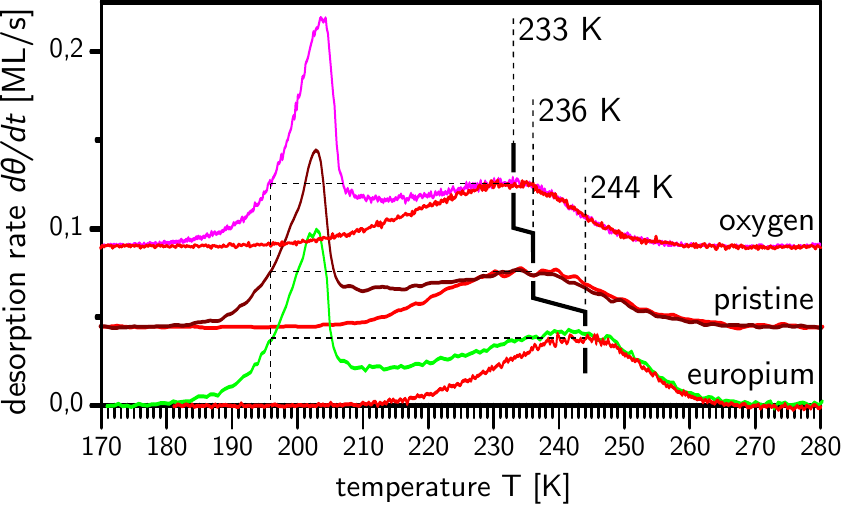}
\caption{\label{fig:tds}(color online)
Mass spectrometer signal of desorbing naphthalene ($m=\unit[128]{u}$) while heating the sample with a ramp rate of $\unit[1]{K/s}$ in dependence of intercalant for the coverages of \unit[1]{ML} and $\approx\unit[(2.3\pm0.2)]{ML}$ each.}
\end{figure}

In order to quantify the difference in $E_b$ observed in STM, we investigated thermal desorption of naphthalene from homogeneous Gr layers, \emph{i.e.}, which are either entirely pristine or entirely Eu p($2\times 2$)-intercalated. 

\newcommand {\fns}[1]{{\footnotesize #1}}
\begin{table}[b]
\begin{threeparttable}
\begin{ruledtabular}
\begin{tabular}{lrrr|rrrr}
            &\multicolumn{3}{c}{experiment}
            &\multicolumn{4}{c}{theory}               \\
            &\multicolumn{1}{c}{$E_D^\mathrm{exp}$} 
            &\multicolumn{1}{c}{$T_{max}$} 
            & $E_b^\mathrm{exp}$
            &$E_b^\mathrm{tot}$ & $E_b^\mathrm{DFT}$ & $E_b^\mathrm{NL}$ 
            &\multicolumn{1}{c}{$E_D^\mathrm{theor}$} \\
            \vspace{0.1cm}
            & [eV]$^{\phantom{\mathrm{a}}}$ & [K]$^{\phantom{\mathrm{a}}}$  & [eV]
            & \multicolumn{4}{c}{[eV]} \\
\cline{2-8}	
Gr/O/Ir	\phantom{$^A$}	
			& \fns{$+0.68$}\tnote{a}
			& \fns{$233^{\phantom{\mathrm{a}}}$}
			& \fns{$-0.808$}
			& \fns{$-0.806$}
			& \fns{$0.366$}
			& \fns{$-1.172$}
			& \fns{$+0.88^{\phantom{\mathrm{b}}}$}
			\\
Gr/Ir		& \fns{$+0.10$}\tnote{b}
			& \fns{$236^{\phantom{\mathrm{a}}}$}
			& \fns{$-0.819$}
			& \fns{$-0.852$} 
			& \fns{$0.556$} 
			& \fns{$-1.408$} 
			& \fns{$+0.15^{\phantom{\mathrm{b}}}$}
			\\
Gr/Eu/Ir	& \fns{$-1.36$}\tnote{c} 
			& \fns{$244^{\phantom{\mathrm{a}}}$}  
			& \fns{$-0.848$}
			& \fns{$-0.865$} 
			& \fns{$0.736$} 
			& \fns{$-1.601$}  
			& \fns{$-1.21^{\phantom{\mathrm{b}}}$}
			\\		
\end{tabular}
\end{ruledtabular}
\begin{tablenotes} %
\item[a] Ref.~\cite{PhysRevB.89.155435}~(also Refs.~\cite{ACSNano.6.9551}~and~\cite{Ulstrup14})
	\item[b]\,Ref.~\cite{PhysRevLett.102.056808}
	\item[c] Ref.~\cite{PhysRevB.90.235437}
\end{tablenotes}
\caption{\label{tab:EdvsTmax}
Doping levels $E_D$, desorption peak maximum $T_{max}$ and binding energies 
$E_b$ from experiment and theory. $E_b^\mathrm{exp}$ is given assuming 
$\nu = \unit[5\times 10^{16}]{Hz}$ (without error as it acts equally on all 
values).}
\end{threeparttable}
\end{table}

Figure~\ref{fig:tds} shows the desorption traces for naphthalene coverages of 1~ML and $\unit[(2.3\pm0.2)]{ML}$. The low-temperature peak corresponds to desorption from the naphthalene multilayer and is zero-order as seen in the exponential leading edge followed by a sharp drop in intensity, while the high temperature peak is almost symmetric, indicative of first-order desorption. 
For pristine graphene, the maximum of the monolayer peak is at $T_{max}=\unit[236]{K}$. These findings agree in the limits of error with those of Zacharias et al.\cite{PhysRevB.69.155406} for naphthalene desorption from graphite, a hint that the influence of the Ir substrate on the binding of naphthalene to Gr is negligible.

Upon intercalation of Eu (2$\times$2) (associated with a $\unit[-1.36]{eV}$ Gr doping level), the monolayer peak is shifted up to \unit[244]{K}, evidence for increased $|E_b|$,
while the position of the multilayer peak is the same as seen in its intercept with the desorption rate at the maximum of the monolayer peak (horizontal dashed lines) at 196~K (left vertical dashed line). As the multilayer peak is not expected to shift due to the diminishing effect of the substrate with increasing molecular film thickness, its presence provides an integrated temperature calibration.

As shown in Fig.~\ref{fig:tds}, by TDS we also investigated O-intercalated Gr, for which the monolayer $T_{max}$ is shifted down to $\unit[233]{K}$. This case would have been difficult to analyze by STM, as the intercalation is not regularly patterned as for the case of Eu. TDS of naphthalene adsorbed to two other Gr intercalation systems [Cs, Eu-($\sqrt{3} \times \sqrt{3}$)] are analyzed in the Supplemental Material \cite{SM}. Altogether, we find a strictly monotonic increase of $|E_b|$ with the Fermi level moving up in the bandstructure from $p$- to $n$-doping. 

To extract $E_b$ from the desorption temperatures, we employed the Redhead method using the previously determined value of the pre-exponential factor $\nu=\unit[5\times 10^{16\pm 2}]{Hz}$ \cite{PhysRevB.69.155406}. The left part of Table~\ref{tab:EdvsTmax} lists the experimentally determined $E_b^\mathrm{exp}$ alongside the doping levels $E^{\mathrm{exp}}_D$, which are given in terms of the position of the Dirac cone relative to the Fermi level as determined in angle-resolved photoemission (ARPES) measurements. The experimental error in $E_b$ is about $\unit[100]{meV}$, and dominated by the error of the pre-exponential factor. However, as the pre-exponential factor acts equally in the calculation of all $E_b$, the errors of the differences
in $E_b$ for the differently doped Gr are dominated by the reproducibility of the temperature measurement, and are therefore much smaller, only about $\unit[4]{meV}$.

\begin{figure}[tb]
	\includegraphics[width=\columnwidth]{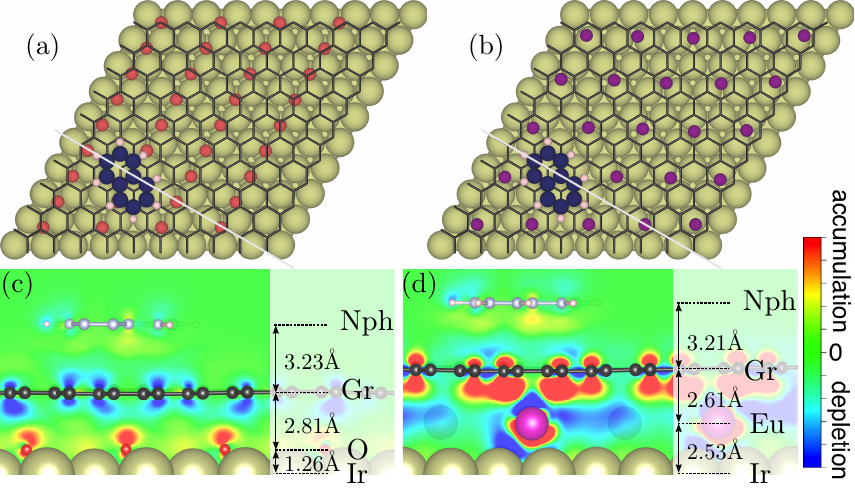}
	\caption{(color online) Top view of relaxed adsorption geometries of 
		naphthalene on (a) Gr/O/Ir(111) and (b) Gr/Eu/Ir(111). In (c) and (d), 
		the corresponding plots of the  charge density difference for each system are also presented.
		Color scale range is $\pm \unit[5\times 10^{-4}]{e/a_0^3}$
		with $a_0$ the Bohr radius.}
	\label{fig:chg_diff}
\end{figure}

In order to better understand the origin of the observed change in $E_b$ upon intercalation,
we have conducted DFT calculations of the naphtalene molecule on non-intercalated as well as O- and Eu-intercalated Gr.
Figure~\ref{fig:chg_diff} (a) and (b) show the geometry of the calculations for the intercalated systems, while the right part of Table~\ref{tab:EdvsTmax} provides the calculated values for doping level $E^\mathrm{theor}_D$ and binding energy $E^\mathrm{tot}_b$.

The calculated doping level $E^\mathrm{theor}_D$ has been determined from the shift of Gr's Dirac point against the Fermi level and should thus be directly comparable to the value extracted from ARPES. Indeed, we find that the calculated doping levels $E^\mathrm{theor}_D$ are in reasonable agreement with the experimentally determined $E_D^\mathrm{exp}$, in particular confirming $n$-doping of Gr by Eu and $p$-doping of Gr by O intercalation.
Qualitatively, the doping is also seen in the charge density difference plots 
in Fig.~\ref{fig:chg_diff}~(c) and (d), which give a direct image of charge accumulation (for Eu) and depletion (for O) on the Gr site. These plots also 
visualize that charge transfer between molecule and Gr is absent (a Bader 
analysis \cite{book_Bader} yields a charge transfer of less 
than 0.03 electrons), which is a precondition for naphthalene adsorbed to Gr being a model for vdW interaction even when Gr is supported on a substrate and doped by intercalation.

Next, we look at the theoretical binding energy, defined as $E_b^\mathrm{tot}=E_\mathrm{sys}^\mathrm{tot}-(E_\mathrm{surf}^\mathrm{tot}+
E_\mathrm{molec}^\mathrm{tot})$, where $E_\mathrm{sys}^\mathrm{tot}$ is the 
total energy of the molecule-surface system, $E_\mathrm{surf}^\mathrm{tot}$ 
denotes the total energy of the surface (i.e., the Gr/Ir(111) and Gr/Eu/Ir(111) 
systems) and $E_\mathrm{molec}^\mathrm{tot}$ corresponds to the total energy 
of the naphthalene molecule in the gas phase.
As obvious from Table~\ref{tab:EdvsTmax}, the calculations reproduce the experimentally observed increase in $|E_b|$ rather well when 
going from Gr/O/Ir(111) via Gr/Ir(111) to naphthalene/Gr/Eu/Ir(111).
Yet, considering absolute numbers, two points have to be noted: Firstly, while Gr is essentially freestanding when intercalated with O or Eu \cite{ACSNano.6.9551, PhysRevB.90.235437}, Gr on bare Ir(111) is slightly hybridized with its substrate in specific areas of its moiré \cite{PhysRevLett.107.036101}. This induces a certain modulation of $E_b$ and thus renders the non-intercalated case less reliable.
Secondly, the excellent match of the absolute values (within $<\unit[25]{meV}$) of the binding energies must be considered fortuitous, as $E_b$ extracted from TDS is expected to carry a larger systematic error.
Unaffected by these two caveats, however, is the \emph{difference} in $E_b$ between the O- and Eu-intercalated cases, which matches reasonably well being $\approx \unit[60]{meV}$ in the calculation and $\approx\unit[40]{meV}$ in the experiment.

To gain an insight into the role of the vdW interactions in these
systems, the total binding energy $E_b^\mathrm{tot}$ was decomposed into a 
DFT contribution $E_b^\mathrm{DFT}$ and a non-local (NL) one $E_b^\mathrm{NL}$
(for details, see Supplemental Material \cite{SM}) as seen in Table~\ref{tab:EdvsTmax}.
The data show that
for all systems considered in our study, the DFT contribution $E_b^\mathrm{DFT}$ 
to $E_b^\mathrm{tot}$ is positive while the NL one $E_b^\mathrm{NL}$ is negative 
indicating that the vdW interactions
are indeed the driving force leading to a stable molecular adsorption.
Furthermore, the magnitude of the attractive NL contribution $E_b^\mathrm{NL}$ to the naphthalene binding significantly increases in the series from the Gr/O/Ir(111) ($\unit[1.172]{eV}$) via
the Gr/Ir(111) ($\unit[1.408]{eV}$) to the Gr/Eu/Ir(111) ($\unit[1.601]{eV}$) system.

\begin{figure}[tb]
	\begin{center}
		\includegraphics[width=\columnwidth]{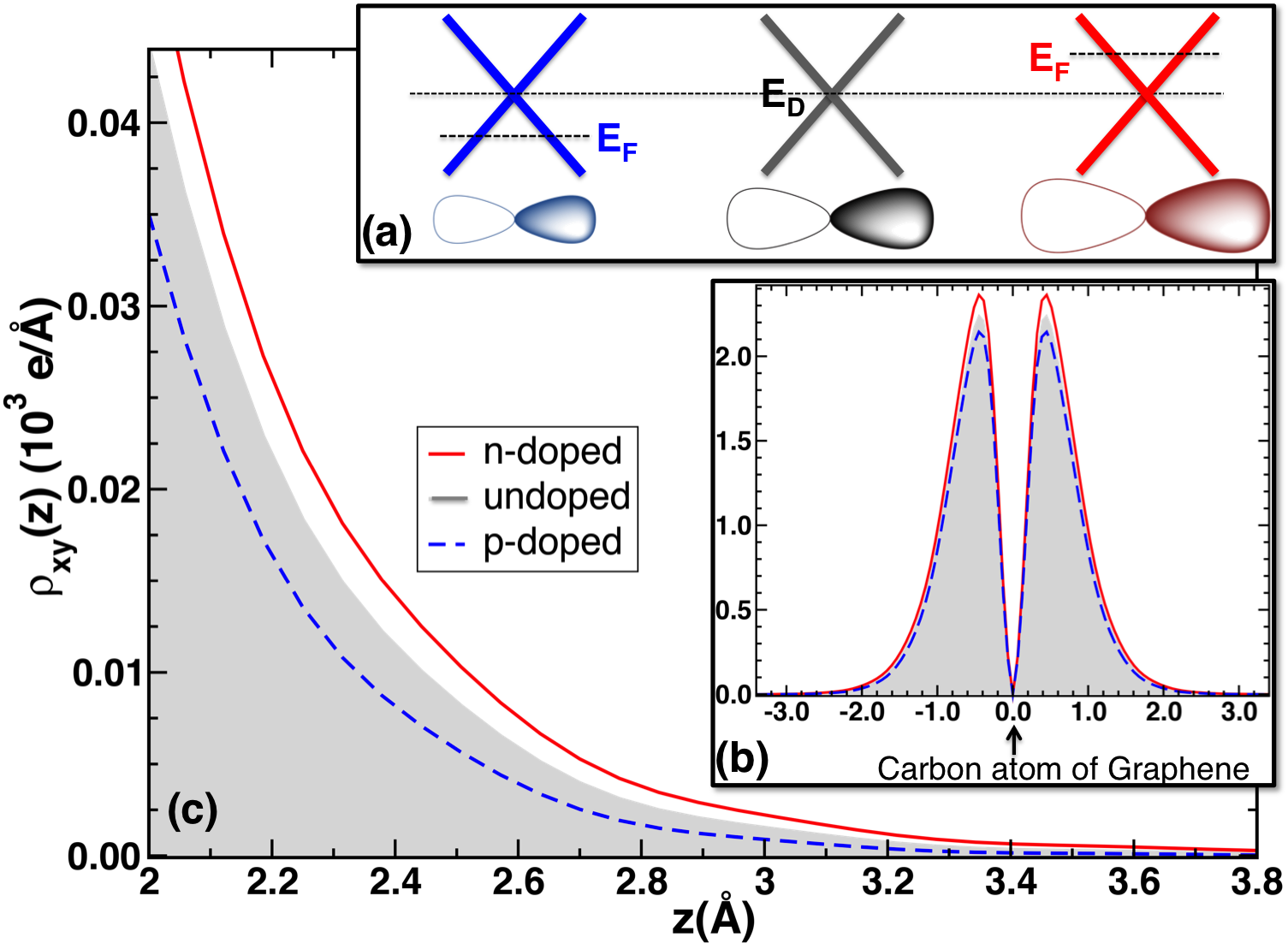}
	\end{center}
	\caption{(color online)
	  (a) Schematic band structure near the Dirac point and representation of the $\pi$ orbital extension for \mbox{$p$-,} undoped and $n$-doped graphene.
		(b) In-plane $xy$-integrated $\pi$-state charge density for (10$\times$10) unit cells of freestanding Gr as a function of $z$-direction.
		(c) Zoom into the low-density tail of~(b).
		}
	\label{fig:theo1}
\end{figure}

To find out the reason for the modification of the vdW binding upon 
intercalation, we have conducted calculations on a freestanding Gr 
sheet of $(10\times 10)$ unit cells. 
It was doped by directly adding or removing 0.05 electrons per C
atom. To keep the calculated unit cell on the whole charge-neutral, a 
balancing background charge had to be introduced. 
Figure~\ref{fig:theo1}~(b) shows the resulting $xy$-integrated charge 
density for all $\pi$ orbitals corresponding to the differently doped 
cases.
As qualitatively illustrated in Fig.~\ref{fig:theo1}\,(c), the $p$- 
or $n$-doping of Gr results in a $\pi$ charge density distribution 
that is less ($p$-doped Gr) or more ($n$-doped Gr) spatially extended 
as compared to that of the undoped Gr. 
A less (more) spatially extended $\pi$ charge density corresponds 
to less (more) polarizable C atoms in the Gr layer and therefore will govern a weaker (stronger) vdW binding of the $\pi$ adsorbates. 
This picture is derived from the behavior observed for the vdW bonded 
noble-gas dimers from He to Kr \cite{PRL92_246401,JCP137_134109}. 
It is backed up quantitatively also by a recent analysis of Berland \emph{et al.} \cite{PhysRevB.87.205421}, who found for benzene on Gr, that most of the non-local energy in the vdW-DF-type functional arises from the extended regions of low charge density.

The increase in the non-local contribution with electron doping in 
Table~\ref{tab:EdvsTmax} is mostly counteracted by the simultaneous 
increase in the repulsive DFT contribution, such that the total binding 
energy $E^{tot}_b$ of the molecule to the surface is modified only 
slightly. Generally in a pure vdW system, the attractive vdW forces have 
to be balanced by a counteracting Pauli repulsion force in any stable 
adsorption geometry. If one approximates the vdW contribution by the non-local correlation,
one might tentatively attribute the increase in the repulsive DFT term to an 
additional Pauli repulsion as Gr's charge density becomes more extended 
due to $n$-doping.

In view of the generality of the mechanism proposed above, we speculate that also substitutional doping or tuning of the Gr Fermi level through a gate electrode will affect the vdW contribution of adsorbate binding in a similar way.

In conclusion, in excellent agreement of experiment and theory we find an increase in the vdW binding energy of naphthalene to Gr by about 5\,\%, when changing from $p$- to $n$-doping in the experimentally accessible range. 
Based on our DFT calculations, the Gr $\pi$ orbitals become spatially more extended with $n$-doping, making the space of low electron density between Gr and the molecule more polarizable, which in turn gives rise to a stronger vdW binding. 
The effect is robust with respect to changing either the intercalant or the molecule and must therefore be assumed to provide a broadly applicable strategy to manipulate van der Waals binding in $\pi$-$\pi$ systems.
	
\section{Acknowledgements}

The computations were performed under the auspices of the VSR at the 
computer JUROPA and the GCS at the high-performance computer JUQUEEN 
operated by the JSC at the Forschungszentrum J\"ulich. 
Financial support from the Volkswagen-Stiftung through the "Optically Controlled Spin Logic"
project (N.\,A. and V.\,C.), from the European Commission through a 
Marie Curie Fellowship (A.\,J.\,M.-G.), from the Institutional Strategy of the University of Cologne
within the German Excellence Initiative (F.\,H.), 
from the 
Ministry of Education, Culture, Sports, Science and Technology of Japan (MEXT) through the ``World Premier International Research Center Initiative'' and the ``Development of Environmental Technology using Nanotechnology'' program (I.\,H.),
and from the DFG Priority Program 
1459 ``Graphene'' within project MI581/20-1 is gratefully acknowledged.

\bibliography{references}

\end{document}


\title{Tuning the van der Waals Interaction of Graphene with Molecules via Doping: Supplemental Material}

\author{Felix Huttmann}\email{huttmann@ph2.uni-koeln.de}
\affiliation{II. Physikalisches Institut, Universit\"{a}t zu K\"{o}ln, 
	Z\"{u}lpicher Stra\ss e 77, 50937 K\"{o}ln, Germany}
\author{Antonio J. Martínez-Galera}
\affiliation{II. Physikalisches Institut, Universit\"{a}t zu K\"{o}ln, 
	Z\"{u}lpicher Stra\ss e 77, 50937 K\"{o}ln, Germany}
\author{Vasile Caciuc}
\affiliation{Peter Grünberg Institute and Institute for Advanced Simulation, 
	Forschungszentrum Jülich, 52428 Jülich, Germany}
\author{Nicolae Atodiresei}\email{n.atodiresei@fz-juelich.de}
\affiliation{Peter Grünberg Institute and Institute for Advanced Simulation, 
	Forschungszentrum Jülich, 52428 Jülich, Germany}
\author{Stefan Schumacher}
\affiliation{II. Physikalisches Institut, Universit\"{a}t zu K\"{o}ln, 
	Z\"{u}lpicher Stra\ss e 77, 50937 K\"{o}ln, Germany}
\author{Sebastian Standop}
\affiliation{II. Physikalisches Institut, Universit\"{a}t zu K\"{o}ln, 
	Z\"{u}lpicher Stra\ss e 77, 50937 K\"{o}ln, Germany}
\author{Ikutaro Hamada}
\affiliation{International Center for Materials Nanoarchitectonics 
	(WPI-MANA) and Global Research Center for Environment and Energy
	based on Nanomaterials Science (GREEN), National Institute for Materials 
	Science, 1-1 Namiki, Tsukuba 305-0044, Japan}
\author{Tim O. Wehling}
\affiliation{Bremen Center for Computational Material Science (BCCMS), 
	Universität Bremen, Am Fallturm 1a, 28359 Bremen, Germany}
\author{Stefan Blügel}
\affiliation{Peter Grünberg Institute and Institute for Advanced Simulation, 
	Forschungszentrum Jülich, 52428 Jülich, Germany}
\author{Thomas Michely}
\affiliation{II. Physikalisches Institut, Universit\"{a}t zu K\"{o}ln, 
	Z\"{u}lpicher Stra\ss e 77, 50937 K\"{o}ln, Germany}
\date{\today}
\maketitle

\newcommand{\beginsupplement}{%
	\setcounter{table}{0}
	\renewcommand{\thetable}{S\arabic{table}}%
	\setcounter{figure}{0}
	\renewcommand{\thefigure}{S\arabic{figure}}%
}
\beginsupplement
This Supplemental Material presents thermal desorption spectroscopy of naphthalene on additional intercalation systems, scanning tunneling microscopy (STM) of benzene on intercalation-patterned graphene (Gr), and a description of the binding energy evaluation from our density functional theory (DFT) calculations.

\section{Full set of desorption traces \\for more intercalation structures}
In the paper, we presented desorption of naphthalene from pristine Gr/Ir(111) and intercalated either with oxygen to saturation or Eu in a p$(2\times2)$ superstructure,
and only two selected coverages of naphtalene for reasons of clarity.

However, we also investigated two other Gr intercalation structures:
(1) Cs adsorbed in a $(\sqrt{3}\times\sqrt{3})$R30$^\circ$ with respect to Ir, prepared by room temperature deposition of Cs from commercial dispensers~\footnote{Cs dispensers obtained from SAES Getters} till saturation, and
(2) Eu adsorbed in a $(\sqrt{3}\times\sqrt{3})$R30$^\circ$ with respect to Gr, prepared by deposition of Eu at \unit[720]{K} until saturation.

\begin{figure}[b]
\begin{center}
\includegraphics[width=0.96\columnwidth]{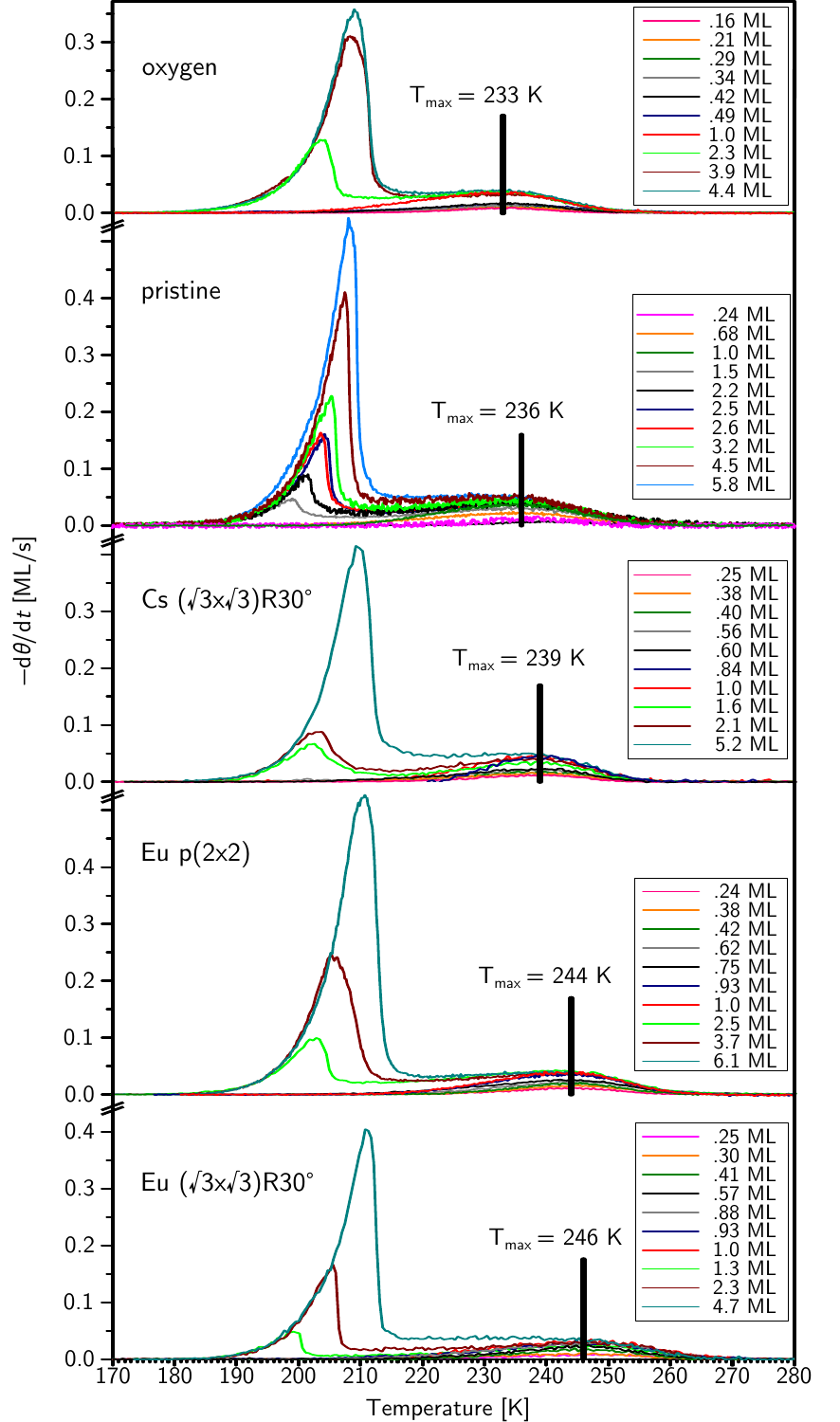}
\end{center}
\caption{\label{fig:alltraces}
	Mass spectrometer signal of desorbing naphthalene ($m=\unit[128]{u}$) while heating the sample with a ramp rate of $\unit[1]{K/s}$ for different intercalants with naphthalene coverages in the range up to a few monolayer (ML).}
\end{figure}

The full set of all our measured desorption traces for the in total five different intercalation systems and many different coverages each are shown in Fig.~\ref{fig:alltraces}. The coverages were determined by integrating the area under the background-substracted spectra.
\renewcommand{\topfraction}{1.0}
\renewcommand{\topfraction}{1.0}
\renewcommand{\textfraction}{0.0}

\begin{figure}[tb]
\begin{center}
\includegraphics[width=0.5\columnwidth]{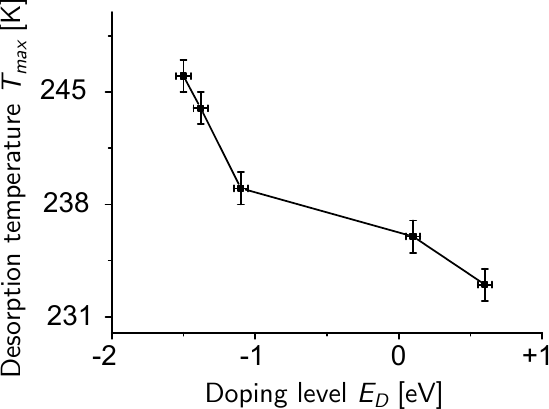}
\end{center}
\caption{\label{fig:TvsED}
Desorption temperature of naphthalene vs. graphene doping. Each data point corresponds to a certain intercalation structure.
}
\end{figure}

As all intercalation structures are associated with a different doping level (see \ref{tab:EdvsTmax} for references), we obtain one point for each in our plot of desorption temperature vs. doping level in Fig.~\ref{fig:TvsED}.
While our expectation of a monotonous increase of naphthalene's binding energy with increasing electron transfer into graphene is confirmed, the effect seems nonlinear and a simple functional dependence cannot be inferred from our present data.

\begin{table}[t]	
\begin{ruledtabular}
\begin{tabular}{lrlllr}
			&	$E_D$&& $T_{max}$&& $E_b^\mathrm{exp}$ \\
			&	[eV] && [K]&&       [eV]\\
\cline{2-6}
Graphite& 0&& 235&\cite{PhysRevB.69.155406}&0.816 \\
Gr/Ir&	$+0.10$&\cite{PhysRevLett.102.056808}& 236&& 0.819\\
Gr/Eu p$(2\times2)$/Ir&		$-1.36$&\cite{PhysRevB.90.235437} & 244&& 0.848\\
Gr/Eu $(\sqrt{3}\times\sqrt{3})$R30$^{\circ}$/Ir&		$-1.43$&\cite{PhysRevB.90.235437}
& 246&& 0.848\\
Gr/Cs $(\sqrt{3}\times\sqrt{3})$R30$^{\circ}$/Ir&		$-1.13$&\cite{PetrovicPhD}& 239&& 0.848\\
Gr/O/Ir&	$+0.68$&\cite{PhysRevB.89.155435}, \cite{ACSNano.6.9551,Ulstrup14}& 233&&	0.808\\
\end{tabular}
\end{ruledtabular}
\caption{\label{tab:EdvsTmax}
Experimental doping levels $E_D$, desorption peak maxima and resulting binding energies $E_b$ for desorption of naphthalene from differently doped Gr.
$E_b^\mathrm{exp}$ is given assuming $\nu = \unit[5\times 10^{16}]{Hz}$ \cite{PhysRevB.69.155406}.
}
\end{table}

\section{Benzene adsorption on intercalation-patterned Gr}
\begin{figure}
	\begin{center}
		\includegraphics[width=1.0\columnwidth]{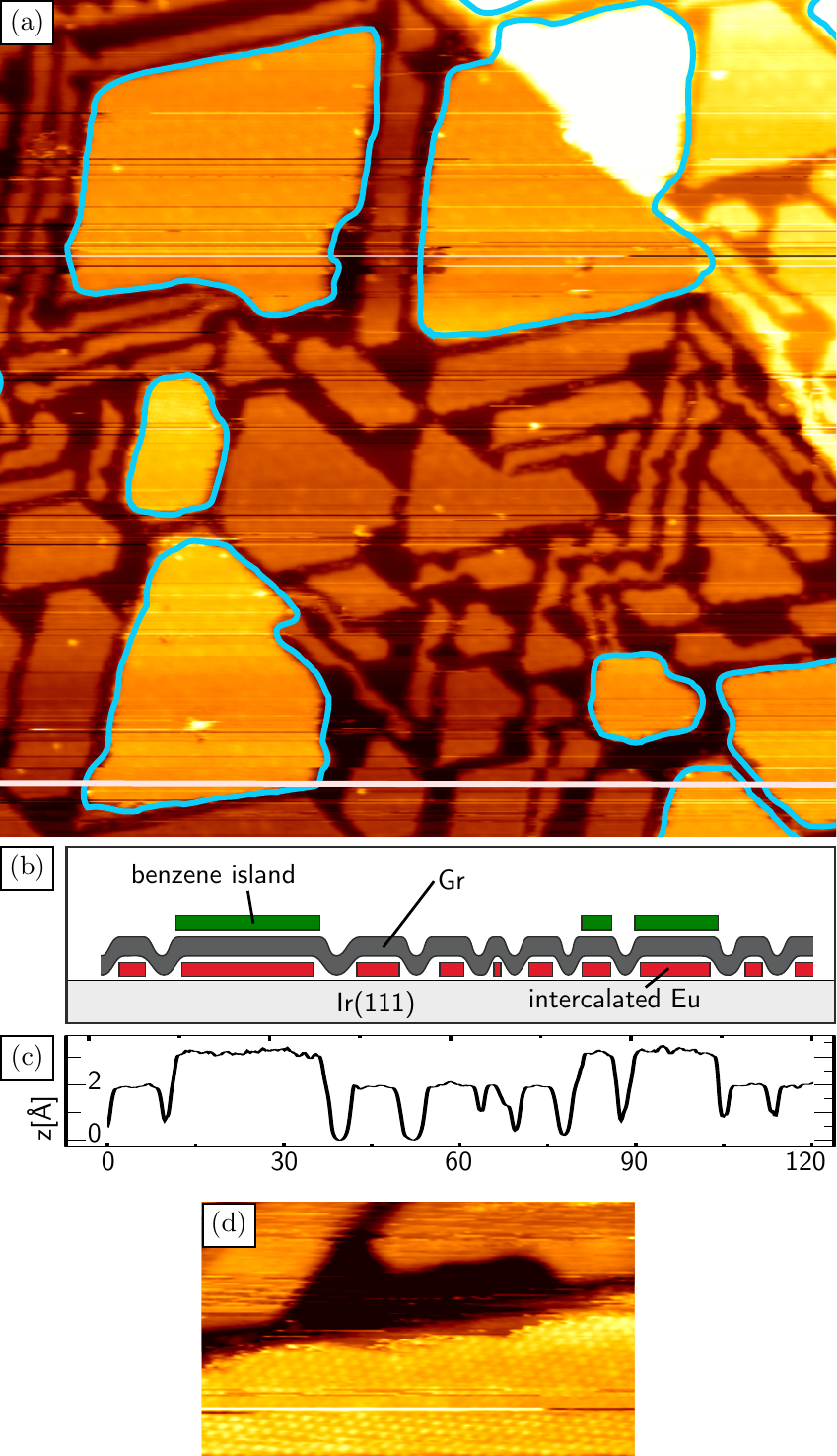}
	\end{center}
	\caption{\label{benzeneislands}
		(a) and (d): STM topographs imaged at $\unit[35]{K}$.
		(a): Partially Eu-intercalated Gr/Ir(111) with submonolayer coverage of benzene ($(\unit[120]{nm})^2$). 
		(b): Schematic cross section corresponding to the line profile in (c).
		(c): Height profile along the white line indicated in (a).		
		(d): Molecular resolution of a benzene island ($\unit[(20\times 12)]{nm^2}$).
		}
\end{figure}

In the paper, we described that on intercalation-patterned Gr, naphthalene islands are preferentially located on the Eu-intercalated areas.
We have repeated the experiment with benzene as a test of generality, and found the same result, as shown in Fig.~\ref{benzeneislands}\,(a)--(d).

However, in the STM topograph in Fig.~\ref{benzeneislands}~(a), another peculiarity is visible: It stands out that benzene islands fill an intercalation island always either completely or not at all. Furthermore, larger intercalation islands are predominantly covered with benzene islands (the three largest islands in the image are all covered), while smaller intercalation islands are less often and thin stripes never covered. 
Our interpretation is that this is a confined Ostwald ripening process:
As larger islands have smaller edge-to-area ratios, they are energetically more favorable. During the annealing, thermal energy is sufficient to make the molecules mobile also over non-intercalated ares. Then, larger benzene islands grow at the expense of smaller ones until the area of preferential binding on the intercalation island is completely filled. 

\section{Evaluation of the binding energy from the DFT calculations}

In DFT \cite{PR136_B864}, the total energy 
of the physical system under consideration is a functional of the electron 
charge density $n(\mathbf{r})$ and can be expressed in the Kohn-Sham 
formalism \cite{PR140_A1133} as 
\begin{equation}
E[n]^\mathrm{tot}=T_s[n]+E_\mathrm{ext}[n]+E_H[n]+E_x[n]+E_c[n].
\label{eq:DFT}
\end{equation}
where $T_S[n]$ describes the kinetic energy, $E_\mathrm{ext}[n]$ is the 
potential energy due to the external potential (\emph{i.e.}, generated by nuclei, 
etc.), $E_H[n]$ corresponds to the potential energy due to the repulsive 
Coulomb electron-electron interaction and $E_x[n]$ and $E_c[n]$ represent 
the exchange and correlation energy, respectively. In particular, in the 
commonly used PBE exchange-correlation functional~\cite{PRL77_3865} the 
correlation part is given by a sum of a LDA and a semi-local term
\begin{equation}
E_c[n]=E_c^\mathrm{LDA}[n]+E_c^\mathrm{SL}[n],
\end{equation}
while in the vdW-DF formalism~\cite{PRL92_246401} the semi-local component 
$E_c^\mathrm{SL}[n]$ of the correlation energy is replaced by a non-local 
term $E_c^\mathrm{NL}[n]$
\begin{equation}
E_c[n]=E_c^\mathrm{LDA}[n]+E_c^\mathrm{NL}[n].
\end{equation}

In consequence, the total energy of the system can be cast into the following 
expression
\begin{equation}
E[n]^\mathrm{tot}=E[n]^\mathrm{DFT}+E_c^\mathrm{NL}[n],
\label{eq:DFT}
\end{equation}
where
\begin{equation}
E[n]^\mathrm{DFT}=T_s[n]+E_\mathrm{ext}[n]+E_H[n]+E_x[n]+
E_c^\mathrm{LDA}[n].
\label{eq:DFT_NL}
\end{equation}
Finally, we employed Eq.\,\ref{eq:DFT_NL} together with the definition 
of the binding energy $E_b^\mathrm{tot}$ of the molecule-surface system 
$E_b^\mathrm{tot}=E_\mathrm{molec+surf}^\mathrm{tot}-
(E_\mathrm{molec}^\mathrm{tot}+E_\mathrm{surf}^\mathrm{tot})$ 
to decompose the $E_b^\mathrm{tot}$ into a DFT and a NL part
\begin{equation}
E_b^\mathrm{tot}=E_b^\mathrm{DFT}+E_b^\mathrm{NL}.
\end{equation}
\vspace{0.3cm}
  
\bibliography{references}